\documentclass[12pt]{iopart}

\usepackage{graphicx}
\usepackage{iopams}
\usepackage{color}
\usepackage{hyperref}

\begin{document}

\title{Evidence for a field-induced Lifshitz transition in the Weyl semimetal CeAlSi}

\author{M. M. Piva$^1$, T. Helm$^2$, J. C. Souza$^{3,1,4}$, K. R. Pakuszewski$^{3,5}$, C. Adriano$^{3,5}$, P. G. Pagliuso$^{3}$, M. Nicklas$^1$\footnote{Author to whom any correspondence should be addressed}}

\address{$^1$Max Planck Institute for Chemical Physics of Solids, N\"{o}thnitzer Str.\ 40, D-01187 Dresden, Germany}
\address{$^2$Hochfeld-Magnetlabor Dresden (HLD-EMFL) and Würzburg-Dresden Cluster of Excellence ct.qmat, Helmholtz-Zentrum Dresden-Rossendorf, 01328 Dresden, Germany}
\address{$^3$Instituto de F\'{\i}sica ``Gleb Wataghin'', UNICAMP, 13083-859, Campinas, SP, Brazil}
\ead{Mario.Piva@cpfs.mpg.de, Michael.Nicklas@cpfs.mpg.de}

\begin{abstract}
The Weyl semimetal CeAlSi crystallizes in the noncentrosymmetric tetragonal space group $I4_1md$ and exhibits ferromagnetic order below 8~K, thereby breaking both spatial inversion and time-reversal symmetries. This unique combination of properties establishes CeAlSi as a model system for studying the interplay between non-trivial topological states and strong electron correlations. In this work, we report observations of Shubnikov–de Haas oscillations in the electrical resistivity under magnetic fields up to 68~T applied parallel to the $[001]$ crystallographic axis. Our measurements reveal an abrupt change in the oscillation frequencies near 14~T, which is indicative of a field-induced Lifshitz transition. Additionally, our results are consistent with the ferromagnetic order bringing the Weyl nodes closer to the Fermi level in CeAlSi. Furthermore, they suggest that the RKKY interaction plays an important role.
\end{abstract}

\vspace{10em}

\address{$^4$Present Address: Department of Condensed Matter Physics, Weizmann Institute of Science, Rehovot, Israel.}
\address{$^5$Present Address: Institut Quantique de l'Université de Sherbrooke, 2500 boul. de l'Université, Sherbrooke (Québec) J1K 2R, Canada.}

\maketitle

\section{Introduction}

Topological phases of matter show great potential for advancing next-generation computing and storage technologies \cite{he2019topological,kumar2020topological}, with Weyl semimetals emerging as promising candidates for realizing such innovations \cite{yang2021chiral}. These materials host massless charge carriers called Weyl fermions, which give rise to exotic electronic phenomena such as extremely large magnetoresistance (MR), the anomalous Hall effect (AHE), Fermi arc surface states, and the chiral anomaly \cite{yan2017topological}. The formation of Weyl nodes -- linear band crossings protected by topology -- in Weyl semimetals requires the breaking of either spatial inversion (SI) symmetry or time-reversal (TR) symmetry, or both.     

Promising candidates include members of the $R$Al$X$ family (where $R =$ is a rare-earth element and $X$ is Ge or Si). These materials, such as the TaAs family of Weyl semimetals \cite{weng2015weyl}, adopt the noncentrosymmetric tetragonal space group $I4_{1}md$, which inherently breaks SI symmetry. Theoretical predictions \cite{ng2021origin} and experimental observations confirm the presence of multiple Weyl nodes in LaAlGe \cite{xu2017discovery} and LaAlSi \cite{kunze2024optical}. Incorporating magnetic rare-earth ions introduces long-range magnetic order, further breaking TR symmetry and positioning these compounds as ideal platforms for exploring Weyl physics. The coexistence of localized magnetic moments also gives rise to intricate phenomena, including the Kondo effect, crystalline electric field (CEF) effects, and the Ruderman–Kittel–Kasuya–Yosida (RKKY) interaction.

An anomalous Hall effect has been reported in PrAlGe$_{1-x}$Si$_{x}$ \cite{yang2020transition} and NdAlGe \cite{kikugawa2024anomalous}, while CeAlGe exhibits a topological Hall effect and a nontrivial magnetic phase accompanied by unconventional magnetoresistance features \cite{piva2023topologicalGe,puphal2020topological,suzuki2019singular}. In the silicon-based analog, a Weyl node-driven magnetic order has been observed in NdAlSi \cite{gaudet2021weyl}, and SmAlSi displays a $\pi$~Berry phase, as confirmed by quantum oscillation (QO) analysis \cite{xu2022shubnikov}. Recent studies have also identified Fermi arcs have in CeAlSi \cite{sakhya2023observation} and NdAlSi \cite{li2023emergence}, while optical reflectivity measurements have confirmed the presence of Weyl fermions in CeAlSi, PrAlSi, NdAlSi, and SmAlSi \cite{kunze2024optical}. 

We focus on CeAlSi, which exhibits a noncollinear ferromagnetic order below 8~K. Previous studies have established that this magnetic order shifts the Weyl nodes closer to the Fermi level \cite{cheng2024tunable}. Additionally, minor shifts in the position of the Fermi level significantly alter the electronic transport response. Hall resistivity measurements demonstrate an evolution from single-band to two-band dominated transport in samples with comparable residual resistivity ratios and stoichiometries \cite{yang2021noncollinear}. These observations highlight the tunability of CeAlSi via external parameters, such as magnetic fields. 

Previous QO studies of CeAlSi have yielded conflicting results. Measurements up to 35~T identified two QO frequencies, of about 40~T and 110~T, which are attributed to different Fermi surface pockets. However, the higher-frequency component exhibits significant sample-to-sample variation from 85~T to 143~T, which has been linked to shifts in the chemical potential relative to the Fermi surface \cite{yang2021noncollinear}. In contrast, our prior low-field measurements ($\leq14$~T) revealed a distinct 20~T QO frequency that disappears at elevated magnetic fields \cite{piva2023topologicalSi}. These discrepancies underscore the intricate electronic structure of CeAlSi and motivate the need for systematic investigations to reconcile these observations.

The sister compound PrAlSi exhibits a remarkable response to high magnetic fields \cite{wu2023field}. An abrupt change in quantum oscillation frequencies occurs near 14.5~T: below this threshold, a dominant 40~T frequency is observed, while higher fields reveal two new frequencies (60~T and 100~T) \cite{wu2023field}. This behaviour has been interpreted as evidence of a field-induced Lifshitz transition. Lifshitz transitions represent topological changes in the Fermi surface topology without symmetry breaking \cite{lifshitz1960anomalies}, and can be triggered by external parameters such as temperature, chemical doping, pressure, or magnetic fields. They are often associated with emergent phenomena, including Van Hove singularities \cite{xu2015lifshitz}, type-0 Lifshitz transitions \cite{Mydeen2017}, unconventional superconductivity \cite{liu2010evidence,li2022elastocaloric,noad2023giant}, and non-trivial topological states \cite{liu2020bond,wu2023field}. Given the structural and electronic similarities between PrAlSi and CeAlSi, the observation of a Lifshitz transition in PrAlSi motivates a targeted investigation of CeAlSi to determine whether magnetic order and strong electron correlations influence the nature of such field-driven topological transformations.

Here, we report measurements of the electrical resistivity on CeAlSi single crystals in pulsed high magnetic fields of up to 68~T, applied parallel to the $[001]$ crystallographic direction. Our results reveal an abrupt change in quantum oscillation frequencies near 14~T, providing evidence for a field-induced Lifshitz transition in CeAlSi. This transition occurs in the ferromagnetic state of CeAlSi, demonstrating how magnetic exchange interactions can tune topological electronic states through mechanisms that differ from the Zeeman-driven process observed in PrAlSi.  Our transport data are consistent with the established positioning of Weyl nodes near the Fermi level in the ferromagnetic state \cite{cheng2024tunable}, and suggest that RKKY-mediated exchange interactions play an important role in the magnetic properties of CeAlSi. These findings provide critical insights into the electronic properties of CeAlSi and offer direct evidence for a field-driven topological transition. The observed behaviour aligns with theoretical predictions and experimental results in related systems, thereby reinforcing the existence of such Lifshitz transitions in magnetic Weyl semimetals.

\section{Methods}

Single crystals of CeAlSi were synthesised using the Al-flux technique, as described in previous work \cite{piva2023topologicalSi}. The tetragonal crystal structure was confirmed via X-ray powder diffraction, while energy dispersive X-ray spectroscopy verified a stoichiometric Ce:Al:Si ratio of 1:1:1. For electrical transport measurements, a four-probe configuration was employed with current aligned parallel to the $[100]$ crystallographic direction using a low-frequency AC resistance bridge. High magnetic field experiments up to 68 T were conducted at the Dresden High Magnetic Field Laboratory using a 70~T pulsed magnet system with a 120~ms pulse duration and a $^{3}$He cryostat insert for temperature control.

\section{Results}

\subsection{Sample characterisation}
\begin{figure}[t!]
\begin{center}
    \includegraphics[width=0.75\linewidth]{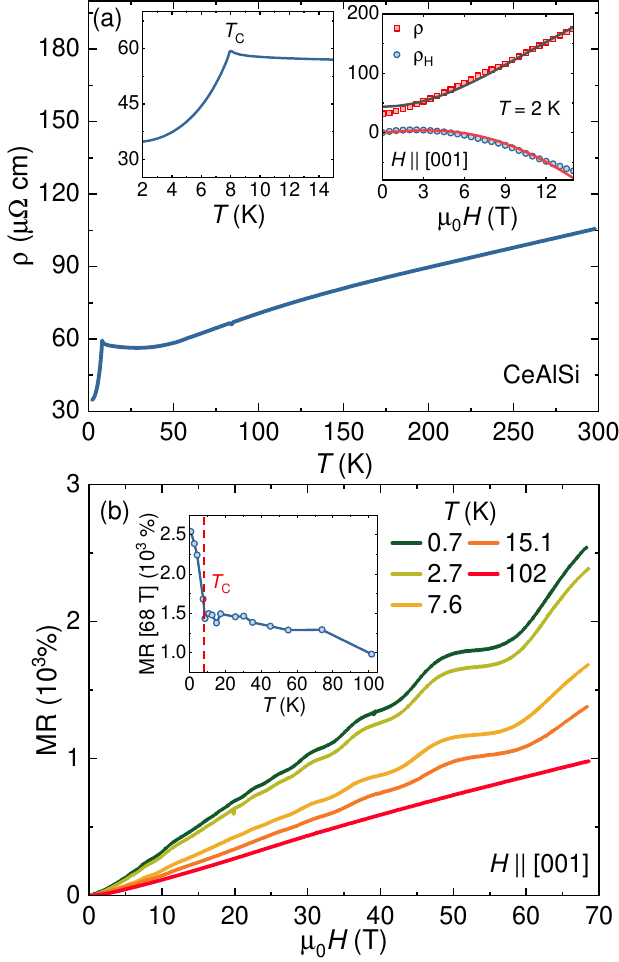}
   	\caption{(a) Electrical resistivity ($\rho$) as a function of temperature for CeAlSi. Left inset: Magnified view of the low temperature region. Right inset: Longitudinal ($\rho$) and Hall resistivity ($\rho_{H}$) as a function of magnetic field applied parallel to $[001]$ at 2~K. The solid lines are two-band model fits used to estimate the charge carrier density. (b) Magnetoresistance, ${\rm MR}=[\rho(H)-\rho(H=0)]/\rho(H=0)$, as a function of magnetic field applied parallel to $[001]$ at different temperatures. Inset: Temperature evolution of the MR value at 68~T. The dashed line indicates $T_{C}$.}
	\label{Rho_MR}
\end{center}
\end{figure}

Figure~\ref{Rho_MR}(a) shows the longitudinal resistivity of CeAlSi as a function of temperature, revealing a sharp kink at 8~K. This feature is attributed to the ferromagnetic phase transition temperature ($T_{C}$), consistent with previous studies \cite{yang2021noncollinear,piva2023topologicalSi}. The field dependence of the longitudinal ($\rho$) and Hall ($\rho_{H}$) resistivities  at 2~K, with a magnetic field up to 14~T applied parallel to the $[001]$ axis, is displayed in the right inset of Fig.~\ref{Rho_MR}(a). These data provide additional insights into the electrical transport properties of the sample. To quantify the charge carrier density, we employed a two-band model described by the following equations:  
\begin{eqnarray}\label{2band}
\rho (H) &=& \frac{1}{e}\frac{(n_{h}\mu_{h} + n_{e}\mu_{e}) + (n_{h}\mu_{e} + n_{e}\mu_{h})\mu_{e}\mu_{h}H^2}{(n_{h}\mu_{h} + n_{e}\mu_{e})^{2}+[(n_{h} - n_{e})\mu_{e}\mu_{h} H]^2} \\
\nonumber \rho_{H} (H) &=& \frac{H}{e}\frac{(n_{h}\mu_{h}^2 - n_{e}\mu_{e}^2) + (n_{h} - n_{e})\mu_{e}^2\mu_{h}^2 H^2}{(n_{h}\mu_{h} + n_{e}\mu_{e})^{2}+[(n_{h} -  n_{e})\mu_{e}\mu_{h} H]^2} 
\end{eqnarray}
where $n$ denotes the hole ($h$) or electron ($e$) carrier density and $\mu$ represents the corresponding mobility. Fitting the experimental data to this model yields $n_e=6.8 \times 10^{19}$~cm$^{-3}$, $n_h=3.3 \times 10^{19}$~cm$^{-3}$, $\mu_{e}=1.1 \times 10^{3}$~cm$^{2}({\rm Vs})^{-1}$, and $\mu_{h}=2.0 \times 10^{3}$~cm$^{2}({\rm Vs})^{-1}$.  

A recent study reported a higher Curie temperature of $T_{C} \approx 10$~K in CeAlSi. This can be attributed to increased defects in non-stoichiometric samples deviating from the 1:1:1 composition \cite{cheng2024tunable}. Such defects may introduce impurity bands, enhancing the charge carrier density and thereby strengthening the RKKY interaction between Ce$^{3+}$ ions, which increases $T_{C}$. This is supported by a linear fit to the Hall resistivity at 2~K in Ref.~\cite{cheng2024tunable}, which yielded $n = 3.4 \times 10^{21}$~cm$^{-3}$. In contrast, our stoichiometric samples exhibit a significantly lower total carrier density of $n = 1.0 \times 10^{20}$~cm$^{-3}$, consistent with previous observations of $T_{C} = 8$~K in systems with a similar charge carrier density \cite{yang2021noncollinear}. The correlation between $T_{C}$ and carrier density is consistent with RKKY-mediated exchange interactions as previously proposed in studies of CeAlSi \cite{yang2021noncollinear,cheng2024tunable}. However, transport measurements alone cannot determine the dominant exchange mechanism. More advanced theoretical frameworks and experimental techniques are required to validate this relationship.

CeAlSi displays a large magnetoresistance, as shown in Fig.~\ref{Rho_MR}(b), for magnetic fields applied parallel to $[001]$. At 0.7~K, the MR reaches approximately $2500\%$ at 68~T, the highest value reported for CeAlSi to date \cite{yang2021noncollinear,cheng2024tunable,alam2023sign}, demonstrating the high quality of the single crystal studied. Notably, as depicted in the inset of Fig.~\ref{Rho_MR}(b), the MR is markedly enhanced below $T_{C}$, indicating a stronger contribution from Weyl fermions to the transport properties in the ferromagnetic state \cite{cheng2024tunable,piva2023topologicalSi}. This enhancement below $T_C$ is consistent with the shift of the Weyl nodes towards the Fermi level in the ferromagnetic state, as
demonstrated by recent angle-resolved photoemission spectroscopy studies \cite{cheng2024tunable}.

\subsection{Shubnikov–de Haas oscillations}

\begin{figure}[!t]
\begin{center}
    \includegraphics[width=0.8\linewidth]{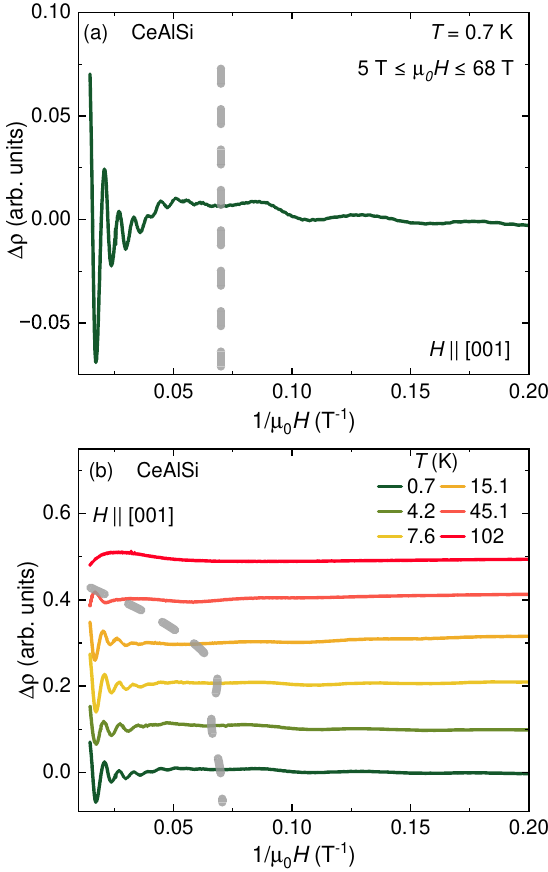}
	\caption{Electrical resistivity ($\Delta \rho$) obtained by subtracting a smooth background (first order polynomial) as a function of inverse magnetic field at  (a) magnified at 0.7~K and (b) at different temperatures to illustrate the temperature evolution. The dashed lines indicate the putative Lifshitz transition.}
	\label{Lifshitz}
\end{center}
\end{figure}

Shubnikov–de Haas oscillations are evident in the MR data shown in Fig.~\ref{Rho_MR}(b). These oscillations are further resolved in Fig.~\ref{Lifshitz}(a), which illustrates the resistivity after subtracting a smooth background ($\Delta \rho$) as a function of the inverse magnetic field applied parallel to the $[001]$ direction at 0.7~K. Low frequency oscillations are observed at low fields between 5 and 14~T (corresponding to $0.07{\rm~T}^{-1}\leqslant (\mu_{0}H)^{-1} \leqslant 0.2 {\rm~T}^{-1}$). Notably, a sharp transition occurs at $H_c \approx 14$~T ($\approx 0.07 {\rm~T}^{-1}$), beyond which oscillations with distinct frequencies emerge. At these fields, the sample is already in a polarised ferromagnetic state, as confirmed by previous studies \cite{yang2021noncollinear,piva2023topologicalSi}. Thus, this transition is not related to changes in the magnetic structure, which have been reported in NdAlSi \cite{gaudet2021weyl}.

Previous resistivity measurements on CeAlSi at high fields did not observe this behaviour \cite{yang2021noncollinear}. This discrepancy likely arises from the extreme sensitivity of CeAlSi to minor variations in the Fermi level, as demonstrated in Ref.~\cite{yang2021noncollinear}. The broadening of the transition makes it difficult to determine the critical field precisely. However, as Fig.~\ref{Lifshitz}(b) shows, the transition shifts to higher fields with increasing temperature. A similar QO frequency shift in the sister compound PrAlSi has been linked to a field-induced Lifshitz transition \cite{wu2023field}. The potential occurrence of such a transition in CeAlSi will be explored in the following discussion.

\begin{figure}[t!]
\begin{center}
    \includegraphics[width=0.75\linewidth]{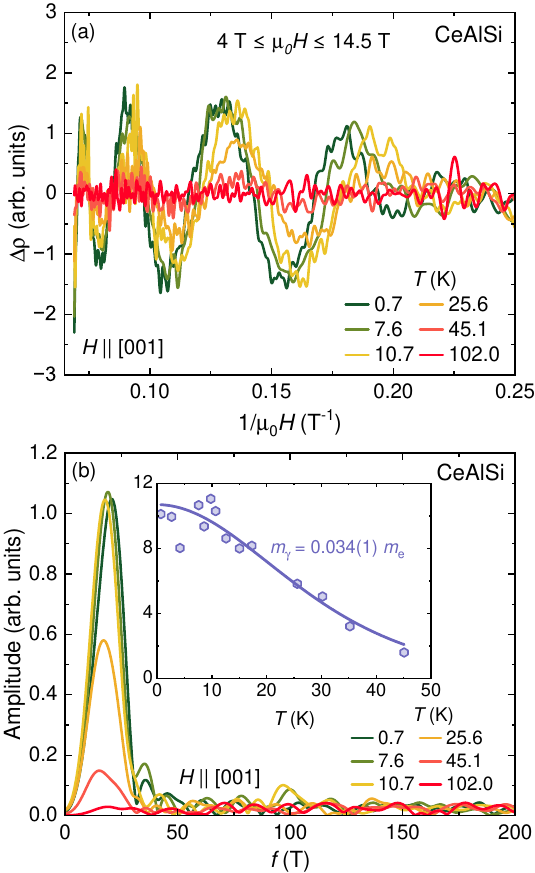}
	\caption{(a) Electrical resistivity  ($\Delta \rho$) obtained by subtracting a smooth background ($7^{\rm th}$ order polynomial) as a function of inverse magnetic field at several temperatures for the low field region ($4 {\rm~T} \leqslant \mu_{0}H \leqslant 14.5 {\rm~T}$). (b) Amplitude of the FFT performed on the curves shown in (a) as a function of frequency at different temperatures. Inset: Amplitudes as a function of temperature and the solid line is a Lifshitz-Kosevich fit.}
	\label{LowField}
\end{center}
\end{figure}

To investigate the field-induced transition, we analyze quantum oscillations in two distinct magnetic field regimes: the low-field region ($4 {\rm~T} \leqslant \mu_{0}H \leqslant 14.5 {\rm~T}$) and the high-field region ($20 {\rm~T} \leqslant \mu_{0}H \leqslant 68 {\rm~T}$). This separation avoids the transition region. Figure~\ref{LowField}(a) displays $\Delta \rho$ as a function of the inverse magnetic field applied along the [001] crystallographic axis at various temperatures for the low-field regime. The oscillation frequency was extracted via fast Fourier transform (FFT), as illustrated in Fig.~\ref{LowField}(b). At low fields, a single frequency $\gamma \approx 20$~T remains dominant up to 45.1~K, exhibiting negligible temperature dependence. The inset in Fig.~\ref{LowField}(b) displays the amplitude of $\gamma$ as a function of temperature. Analysis using the Lifshitz-Kosevich 
 The effective mass ($m^{*}$) was estimated by fitting the FFT amplitude as a function of temperature by the Lifshitz-Kosevich formula \cite{shoenberg2009magnetic}:
\begin{equation}\label{LK}
	A_{T} = \frac{\alpha T m^{*}}{B \sinh (\alpha T m^{*}/B)},
\end{equation}
in which $\alpha=2 \pi^{2} k_{B}/e \hbar \approx 14.69$~T/K, $T$ is the temperature, $B=\mu_0H$ is the magnetic field and $m^{*}$ the effective mass. Note that this represents only the temperature damping factor of the complete LK formula, which we used specifically for the determination of the effective mass. The
model yields an effective cyclotron mass $m_{\gamma} = 0.034(1)m_{e}$, where $m_{e}$ is the free electron mass. This value differs from that reported in Ref.~\cite{piva2023topologicalSi}. Furthermore, the temperature dependence of the $\gamma$-oscillation amplitude is less well defined compared than in Ref.~\cite{piva2023topologicalSi}. For the sample studied here, a small maximum in the amplitude appears near 10~K, but the pronounced suppression of the QO amplitude observed in the ferromagnetically ordered state, observed in Ref.~\cite{piva2023topologicalSi}, is absent. The $\gamma$ frequency has been detected before only in a prior study on samples from the same growth batch \cite{piva2023topologicalSi} and in a recent work by Meena et al.\ \cite{OSC_20T_2025}, but not in other investigations \cite{yang2021noncollinear,cheng2024tunable}. The likely reason for these differences is the high sensitivity of CeAlSi to even minor changes in the position of the Fermi level.

\begin{figure}[t!]
\begin{center}
    \includegraphics[width=0.75\linewidth]{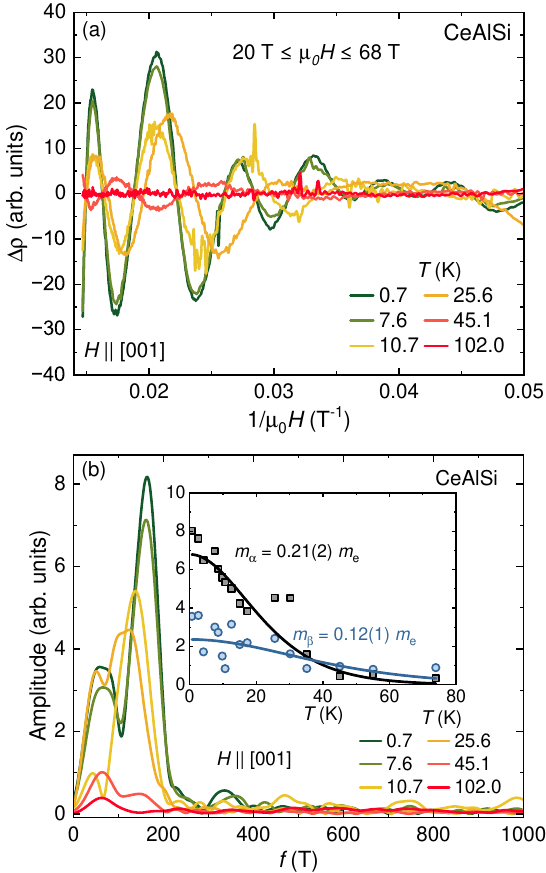}
	\caption{(a) Electrical resistivity ($\Delta \rho$) obtained by subtracting  a smooth background ($5^{\rm th}$ order polynomial) as a function of inverse magnetic field at several temperatures for the high field region ($20 {\rm~T} \leqslant \mu_{0}H \leqslant 68 {\rm~T}$). (b) Amplitude of the FFT performed on the curves shown in (a) as a function of frequency at different temperatures. Inset: Amplitudes as a function of temperature and the solid lines are Lifshitz-Kosevich fits.}
	\label{HighField}
\end{center}
\end{figure}

Next, we analyse the Shubnikov–de Haas oscillations in the high-field region ($20~{\rm T} \leqslant \mu_{0}H \leqslant 68~{\rm T}$). The corresponding $\Delta\rho$ data, plotted as a function of the inverse magnetic field applied parallel to the $[001]$ direction, are shown in Fig.~\ref{HighField}(a). Oscillations persist up to 45.1~K, and fast Fourier transform (FFT) analysis performed analogously to that in the low-field region resolves two distinct frequencies: $\beta \approx 50~{\rm T}$ and $\alpha \approx 150~{\rm T}$ (Fig.~\ref{HighField}(b)). These values are consistent with those obtained in previous high-field measurements on CeAlSi \cite{yang2021noncollinear}. The temperature dependence of the oscillation amplitudes, displayed in the inset of Fig.~\ref{HighField}(b), was fitted using the Lifshitz–Kosevich formula (see Eq.~\ref{LK}). The extracted effective masses, $m_{\beta} = 0.12(1)m_{e}$ and $m_{\alpha} = 0.21(2)m_{e}$, are in agreement with previously reported results \cite{yang2021noncollinear}.  

To ensure consistency, both the low- and high-field datasets were processed in the same way. A digital lock-in amplifier with a third-order low-pass filter and a  of 180~$\mu{\rm s}$ was used, followed by background subtraction via seventh- and fifth-order polynomial fits, respectively. FFT analysis employed a Hamming window and uniform zero-padding across all datasets. This rigorous approach eliminates artefacts from the processing of the data, as the $\alpha$ and $\beta$ frequencies vanish at low fields even with the same treatment. Notably, the $\gamma$ frequency observed at low fields disappears after fifth-order polynomial subtraction in the high-field regime, confirming that it is not an artefact of overfitting. The abrupt emergence of the $\alpha$ and $\beta$ oscillations at 14~T, concomitant with the disappearance of the $\gamma$ frequency, provides compelling evidence for a field-induced Lifshitz transition in CeAlSi.

Our quantum oscillation measurements reveal a dominant frequency of $\gamma = 20~\mathrm{T}$ in CeAlSi that persists from the paramagnetic to the ferromagnetic state without significant change. This is in line with previous observations~\cite{piva2023topologicalSi,OSC_20T_2025}. The calculated cross-sectional area for this frequency, given by $A = 2\pi eF/\hbar = 0.00195~\mathrm{\AA}^{-2}$, yields a Fermi wave vector of $k_F = \sqrt{A/\pi} = 0.025~\mathrm{\AA}^{-1}$, which is consistent with the experimentally reported range of momentum scales from ARPES measurements in the ferromagnetic state  \cite{cheng2024tunable} as reported before by Meena et al.\ \cite{OSC_20T_2025}. At the field-induced Lifshitz transition ($H_c \approx 14~\mathrm{T}$), the $\gamma$ frequency disappears and is replaced by two new frequencies ($\beta = 50~\mathrm{T}$ and $\alpha = 150~\mathrm{T}$) indicating a topological reconstruction of the Fermi surface. The correspondence between specific frequencies and particular Fermi surface pockets in CeAlSi remains challenging to establish definitively.

\section{Discussion}

In order to identify the origin of the abrupt frequency shift at 14 T, we evaluate potential mechanisms, with a particular focus on the sudden change in QO frequencies. This behaviour mirrors the field-induced Lifshitz transition observed in the isostructural compound PrAlSi \cite{wu2023field}), in which a topological Fermi surface reconstruction occurs at a critical magnetic field.

The abrupt emergence of the $\alpha$ and $\beta$ oscillations at 14 T, concomitant with the disappearance of the $\gamma$ frequency, provides compelling evidence for a field-induced Lifshitz transition in CeAlSi. Similar Lifshitz transitions in Weyl semimetals are often associated with emergent phenomena, including Van Hove singularities \cite{xu2015lifshitz}, unconventional superconductivity \cite{liu2010evidence,li2022elastocaloric}, and non-trivial topological states \cite{wu2023field,liu2020bond}.

In Dirac and Weyl semimetals, such as HfTe$_5$ \cite{wang2020approaching,galeski2022signatures,piva2024importance}, ZrTe$_5$ \cite{tang2019three,galeski2021origin}, and TaAs \cite{ramshaw2018quantum}, Lifshitz transitions often occur when the magnetic field drives the system into the quantum limit (where only the lowest Landau level is occupied). However, for CeAlSi the cyclotron energy $ E_c \approx 50 $ meV (calculated as $ E_c = \hbar \omega_c = \hbar eB/m^* $) is significantly smaller than the Fermi energy $ E_F \approx 1800 $ meV ($ E_F = \hbar^2 (3\pi^2 n)^{2/3}/2m^* $). This disparity indicates that the quantum limit is not reached, thus ruling out this mechanism. For comparison, PrAlSi exhibits a similar $ E_c/E_F $ ratio, yet still undergoes a Lifshitz transition \cite{wu2023field}, suggesting that quantum limit effects are not a prerequisite in this material family.

Magnetic fields could alter CEF levels, inducing Fermi surface changes as observed in CeRhIn$_5$ \cite{rosa2019enhanced,lesseux2020orbitally}. However, this scenario is implausible for CeAlSi: the first CEF excitation lies at 23~meV (270~K) above the ground state \cite{yang2021noncollinear,souza2022magnetic}, requiring fields exceeding 100~T (via Zeeman splitting $ \mu_B B \sim k_B \Delta_{\rm CEF} $) to perturb the CEF hierarchy. Such fields are far beyond the 14~T threshold observed here.

Field-induced Lifshitz transitions are common in heavy-fermion systems due to renormalised bands and small effective Fermi energies \cite{aoki1993transition,kozlova2005magnetic,bastien2016lifshitz,pfau2017cascade}. These transitions arise from Zeeman-driven shifts in band structure. However, CeAlSi exhibits negligible $f$-electron hybridisation with conduction bands, as evidenced by its small Sommerfeld coefficient $\gamma_s\approx 50$~mJmol$^{-1}$K$^{-2}$ \cite{souza2022magnetic} and low effective mass $m^*\approx0.034(1) m_e $. This contrasts with heavy-fermion systems (e.g., YbRh$_2$Si$_2$, $\gamma_s\sim1000$~mJmol$^{-1}$K$^{-2} $) \cite{pfau2013interplay}, where strong correlations enable field-tunable Fermi surfaces.

A sudden Fermi surface reconstruction could also arise from magnetic breakdown, where tunnelling between orbits creates new QO frequencies. However, the exclusive observation of the $\gamma$ frequency at low fields argues against this mechanism. Magnetic breakdown typically generates additional frequencies at lower fields due to orbit coupling, rather than a single frequency.

In non-magnetic Weyl semimetals such as TaAs, TaP, or LaAlSi, Zeeman splitting can shift Weyl nodes in magnetic fields, but this arises from intrinsic spin-orbit coupling \cite{xu2015lifshitz,ramshaw2018quantum}. Similarly, in magnetic $R$AlSi compounds such as  PrAlSi, Wu et al.\ \cite{wu2023field} observed a field-induced Lifshitz transition at 14.5 T, which they attributed to Zeeman-driven shifts of Weyl nodes across the Fermi level. However, in CeAlSi, the negligible $ f $-electron hybridisation and rigid-band behaviour (predicted by DFT calculations \cite{cheng2024tunable}) rule out Zeeman-induced band renormalisation as the origin of the frequency change. For instance, the negligible temperature dependence of the $\gamma$ frequency ($ m_\gamma = 0.034 m_e$) is consistent with a Fermi surface stabilised by Weyl nodes near the Fermi level, which are insensitive to thermal or Zeeman-driven perturbations. This distinguishes CeAlSi from PrAlSi, where stronger exchange interactions and higher effective masses enable Zeeman-splitting-driven Lifshitz transitions \cite{wu2023field}.

While alternative explanations cannot adequately account for the abrupt change in QO frequencies at 14 T, the data strongly support the occurrence of a field-induced Lifshitz transition in CeAlSi. This finding highlights the distinct role of magnetic exchange interactions in shaping topological Fermi surfaces and differentiates CeAlSi from its isostructural counterpart PrAlSi. In PrAlSi, Zeeman-driven band renormalisation facilitates the Lifshitz transition \cite{wu2023field}, whereas the negligible $f$-electron hybridisation and rigid-band behaviour \cite{cheng2024tunable} observed in CeAlSi suggest a distinct mechanism tied to the interplay between ferromagnetic order and Weyl node topology. The observed transition demonstrates how magnetic exchange can tune topological states in Weyl semimetals and offers a way to control the reconstruction of the Fermi surface via external fields. These results are consistent with theoretical models of Lifshitz transitions in correlated systems and emphasise the significance of material-specific properties, such as carrier density, exchange interactions, and band structure, in governing such phenomena.  

\section{Conclusion}

We reported evidence for a field-induced Lifshitz transition in the Weyl semimetal CeAlSi through quantum oscillation measurements in magnetic fields of up to 68~T. The abrupt shift from a single low-field frequency ($\gamma = 20$~T) to two high-field frequencies ($\beta = 50$~T, $\alpha = 150$~T) at $H_c \approx 14$~T strongly suggests a topological Fermi surface reconstruction. This transition occurs in the ferromagnetically ordered state of CeAlSi, demonstrating the interplay between magnetic exchange interactions and Weyl fermion physics. The absence of Zeeman-driven band renormalisation or heavy-fermion effects emphasises the distinct mechanism that governs this transition in CeAlSi, compared to other magnetic Weyl semimetals. Further studies, including angle-dependent quantum oscillations and field-dependent band-structure calculations, are required to fully elucidate the topological nature of this transition and its sensitivity to Fermi-level tuning.

\section*{Data availability statement}
Data that underpin the findings of this study are available at Edmond - the open research data repository of the Max Planck Society at https://doi.org/10.17617/3.C3CCTV.

\section*{Acknowledgements}

We thank U.\ Burkhardt for carrying out energy and wavelength dispersive X-ray analysis of the samples. This project has received funding from the European Union’s Horizon 2020 research and innovation programme under the Marie Sk\l{}odowska-Curie grant agreement No 101019024. This work was supported by HLD-HZDR, member of the European Magnetic Field Laboratory (EMFL). We also acknowledge support from FAPESP (SP-Brazil) grant No. 2020/12283-0, 2018/11364-7 and 2017/10581-1, CNPq (Grants \# 405408/2023-4 and 311783/2021-0), CAPES and FINEP-Brazil.

\section*{References}
\bibliographystyle{iopart-num}
\bibliography{CeAlSi_HF}

\providecommand{\newblock}{}
\begin{thebibliography}{10}
\expandafter\ifx\csname url\endcsname\relax
  \def\url#1{{\tt #1}}\fi
\expandafter\ifx\csname urlprefix\endcsname\relax\def\urlprefix{URL }\fi
\providecommand{\eprint}[2][]{\url{#2}}

\bibitem{he2019topological}
He M~Y, Sun H~M and He Q~L 2019 {\em Front. Phys.\/} {\bf 14} 43401
  \urlprefix\url{https://link.springer.com/article/10.1007/s11467-019-0893-4}

\bibitem{kumar2020topological}
Kumar N, Guin S~N, Manna K, Shekhar C and Felser C 2020 {\em Chem. Rev.\/} {\bf
  121} 2780--2815
  \urlprefix\url{https://pubs.acs.org/doi/10.1021/acs.chemrev.0c00732}

\bibitem{yang2021chiral}
Yang S~H, Naaman R, Paltiel Y and Parkin S~S 2021 {\em Nature Reviews
  Physics\/} {\bf 3} 328--343
  \urlprefix\url{https://www.nature.com/articles/s42254-021-00302-9}

\bibitem{yan2017topological}
Yan B and Felser C 2017 {\em Annu. Rev. Condens. Matter Phys.\/} {\bf 8}
  337--354
  \urlprefix\url{https://doi.org/10.1146/annurev-conmatphys-031016-025458}

\bibitem{weng2015weyl}
Weng H, Fang C, Fang Z, Bernevig B~A and Dai X 2015 {\em Phys. Rev. X\/} {\bf
  5}(1) 011029
  \urlprefix\url{https://link.aps.org/doi/10.1103/PhysRevX.5.011029}

\bibitem{ng2021origin}
Ng T, Luo Y, Yuan J, Wu Y, Yang H and Shen L 2021 {\em Phys. Rev. B\/} {\bf
  104}(1) 014412
  \urlprefix\url{https://link.aps.org/doi/10.1103/PhysRevB.104.014412}

\bibitem{xu2017discovery}
Xu S~Y, Alidoust N, Chang G, Lu H, Singh B, Belopolski I, Sanchez D~S, Zhang X,
  Bian G, Zheng H, Husanu M~A, Bian Y, Huang S~M, Hsu C~H, Chang T~R, Jeng H~T,
  Bansil A, Neupert T, Strocov V~N, Lin H, Jia S and Hasan M~Z 2017 {\em Sci.
  Adv.\/} {\bf 3} e1603266
  \urlprefix\url{https://www.science.org/doi/abs/10.1126/sciadv.1603266}

\bibitem{kunze2024optical}
Kunze J, K\"opf M, Cao W, Qi Y and Kuntscher C~A 2024 {\em Phys. Rev. B\/} {\bf
  109}(19) 195130
  \urlprefix\url{https://link.aps.org/doi/10.1103/PhysRevB.109.195130}

\bibitem{yang2020transition}
Yang H~Y, Singh B, Lu B, Huang C~Y, Bahrami F, Chiu W~C, Graf D, Huang S~M,
  Wang B, Lin H, Torchinsky D, Bansil A and Tafti F 2020 {\em APL Mater.\/}
  {\bf 8} 011111 \urlprefix\url{https://doi.org/10.1063/1.5132958}

\bibitem{kikugawa2024anomalous}
Kikugawa N, Uji S and Terashima T 2024 {\em Phys. Rev. B\/} {\bf 109}(3) 035143
  \urlprefix\url{https://link.aps.org/doi/10.1103/PhysRevB.109.035143}

\bibitem{piva2023topologicalGe}
Piva M~M, Souza J~C, Lombardi G~A, Pakuszewski K~R, Adriano C, Pagliuso P~G and
  Nicklas M 2023 {\em Phys. Rev. Mater.\/} {\bf 7}(7) 074204
  \urlprefix\url{https://link.aps.org/doi/10.1103/PhysRevMaterials.7.074204}

\bibitem{puphal2020topological}
Puphal P, Pomjakushin V, Kanazawa N, Ukleev V, Gawryluk D~J, Ma J, Naamneh M,
  Plumb N~C, Keller L, Cubitt R, Pomjakushina E and White J~S 2020 {\em Phys.
  Rev. Lett.\/} {\bf 124}(1) 017202
  \urlprefix\url{https://link.aps.org/doi/10.1103/PhysRevLett.124.017202}

\bibitem{suzuki2019singular}
Suzuki T, Savary L, Liu J~P, Lynn J~W, Balents L and Checkelsky J~G 2019 {\em
  Science\/} {\bf 365} 377--381
  \urlprefix\url{https://www.science.org/doi/abs/10.1126/science.aat0348}

\bibitem{gaudet2021weyl}
Gaudet J, Yang H~Y, Baidya S, Lu B, Xu G, Zhao Y, Rodriguez-Rivera J~A,
  Hoffmann C~M, Graf D~E, Torchinsky D~H, Nikoli{\'c} P, Vanderbilt D, Tafti F
  and Broholm C~L 2021 {\em Nat. Mater.\/} {\bf 20} 1650--1656
  \urlprefix\url{https://www.nature.com/articles/s41563-021-01062-8}

\bibitem{xu2022shubnikov}
Xu L, Niu H, Bai Y, Zhu H, Yuan S, He X, Han Y, Zhao L, Yang Y, Xia Z, Liang Q
  and Tian Z 2022 {\em J. Phys.: Cond. Matter\/} {\bf 34} 485701
  \urlprefix\url{https://iopscience.iop.org/article/10.1088/1361-648X/ac987a}

\bibitem{sakhya2023observation}
Sakhya A~P, Huang C~Y, Dhakal G, Gao X~J, Regmi S, Wang B, Wen W, He R~H, Yao
  X, Smith R, Sprague M, Gao S, Singh B, Lin H, Xu S~Y, Tafti F, Bansil A and
  Neupane M 2023 {\em Phys. Rev. Mater.\/} {\bf 7}(5) L051202
  \urlprefix\url{https://link.aps.org/doi/10.1103/PhysRevMaterials.7.L051202}

\bibitem{li2023emergence}
Li C, Zhang J, Wang Y, Liu H, Guo Q, Rienks E, Chen W, Bertran F, Yang H,
  Phuyal D, Fedderwitz H, Thiagarajan B, Dendzik M, Berntsen M~H, Shi Y, Xiang
  T and Tjernberg O 2023 {\em Nat. Commun.\/} {\bf 14} 7185
  \urlprefix\url{https://www.nature.com/articles/s41467-023-42996-8}

\bibitem{cheng2024tunable}
Cheng E, Yan L, Shi X, Lou R, Fedorov A, Behnami M, Yuan J, Yang P, Wang B,
  Cheng J~G, Xu Y, Xu Y, Xia W, Pavlovskii N, Peets D~C, Zhao W, Wan Y,
  Burkhardt U, Guo Y, Li S, Felser C, Yang W and B{\"u}chner B 2024 {\em Nat.
  Commun.\/} {\bf 15} 1467
  \urlprefix\url{https://www.nature.com/articles/s41467-024-45658-5}

\bibitem{yang2021noncollinear}
Yang H~Y, Singh B, Gaudet J, Lu B, Huang C~Y, Chiu W~C, Huang S~M, Wang B,
  Bahrami F, Xu B, Franklin J, Sochnikov I, Graf D~E, Xu G, Zhao Y, Hoffman
  C~M, Lin H, Torchinsky D~H, Broholm C~L, Bansil A and Tafti F 2021 {\em Phys.
  Rev. B\/} {\bf 103}(11) 115143
  \urlprefix\url{https://link.aps.org/doi/10.1103/PhysRevB.103.115143}

\bibitem{piva2023topologicalSi}
Piva M~M, Souza J~C, Brousseau-Couture V, Sorn S, Pakuszewski K~R, John J~K,
  Adriano C, C\^ot\'e M, Pagliuso P~G, Paramekanti A and Nicklas M 2023 {\em
  Phys. Rev. Res.\/} {\bf 5}(1) 013068
  \urlprefix\url{https://link.aps.org/doi/10.1103/PhysRevResearch.5.013068}

\bibitem{wu2023field}
Wu L, Chi S, Zuo H, Xu G, Zhao L, Luo Y and Zhu Z 2023 {\em npj Quantum
  Mater.\/} {\bf 8} 4
  \urlprefix\url{https://www.nature.com/articles/s41535-023-00537-y}

\bibitem{lifshitz1960anomalies}
Lifshitz I {\em et~al.\/} 1960 {\em Sov. Phys. JETP\/} {\bf 11} 1130--1135

\bibitem{xu2015lifshitz}
Xu S~Y, Liu C, Belopolski I, Kushwaha S~K, Sankar R, Krizan J~W, Chang T~R,
  Polley C~M, Adell J, Balasubramanian T, Miyamoto K, Alidoust N, Bian G,
  Neupane M, Jeng H~T, Huang C~Y, Tsai W~F, Okuda T, Bansil A, Chou F~C, Cava
  R~J, Lin H and Hasan M~Z 2015 {\em Phys. Rev. B\/} {\bf 92}(7) 075115
  \urlprefix\url{https://link.aps.org/doi/10.1103/PhysRevB.92.075115}

\bibitem{Mydeen2017}
Mydeen K, Kasinathan D, Koz C, Rossler S, Rossler U~K, Hanfland M, Tsirlin A~A,
  Schwarz U, Wirth S, Rosner H and Nicklas M 2017 {\em Phys. Rev. Lett.\/} {\bf
  119} 227003 \urlprefix\url{https://www.ncbi.nlm.nih.gov/pubmed/29286759}

\bibitem{liu2010evidence}
Liu C, Kondo T, Fernandes R~M, Palczewski A~D, Mun E~D, Ni N, Thaler A~N,
  Bostwick A, Rotenberg E, Schmalian J, L~Bud’ko S, Canfield P~C and Kaminski
  A 2010 {\em Nat. Phys.\/} {\bf 6} 419--423
  \urlprefix\url{https://www.nature.com/articles/nphys1656}

\bibitem{li2022elastocaloric}
Li Y~S, Garst M, Schmalian J, Ghosh S, Kikugawa N, Sokolov D~A, Hicks C~W,
  Jerzembeck F, Ikeda M~S, Hu Z, Ramshaw B~J, Rost A~W, Nicklas M and Mackenzie
  A~P 2022 {\em Nature\/} {\bf 607} 276--280
  \urlprefix\url{https://www.nature.com/articles/s41586-022-04820-z}

\bibitem{noad2023giant}
Noad H~M, Ishida K, Li Y~S, Gati E, Stangier V, Kikugawa N, Sokolov D~A,
  Nicklas M, Kim B, Mazin I~I, Garst M, Schmalian J, Mackenzie A~P and Hicks
  C~W 2023 {\em Science\/} {\bf 382} 447--450
  \urlprefix\url{https://www.science.org/doi/10.1126/science.adf3348}

\bibitem{liu2020bond}
Liu Y, Liu Y~F, Gui X, Xiang C, Zhou H~B, Hsu C~H, Lin H, Chang T~R, Xie W and
  Jia S 2020 {\em Proc. Nat. Acad. Sci. U.S.A\/} {\bf 117} 15517--15523
  \urlprefix\url{https://www.pnas.org/doi/full/10.1073/pnas.1917697117}

\bibitem{alam2023sign}
Alam M~S, Fakhredine A, Ahmad M, Tanwar P~K, Yang H~Y, Tafti F, Cuono G, Islam
  R, Singh B, Lynnyk A, Autieri C and Matusiak M 2023 {\em Phys. Rev. B\/} {\bf
  107}(8) 085102
  \urlprefix\url{https://link.aps.org/doi/10.1103/PhysRevB.107.085102}

\bibitem{shoenberg2009magnetic}
Shoenberg D 2009 {\em Magnetic oscillations in metals\/} (Cambridge university
  press)

\bibitem{OSC_20T_2025}
Meena P, Mudgal M, Bagga S, Tiwari V~K, Malik V~K, Yadav C~S and Nayak J 2025
  {\em Phys. Rev. Mater.\/} {\bf 9}(4) 044202
  \urlprefix\url{https://link.aps.org/doi/10.1103/PhysRevMaterials.9.044202}

\bibitem{wang2020approaching}
Wang P, Ren Y, Tang F, Wang P, Hou T, Zeng H, Zhang L and Qiao Z 2020 {\em
  Phys. Rev. B\/} {\bf 101} 161201
  \urlprefix\url{https://link.aps.org/doi/10.1103/PhysRevB.101.161201}

\bibitem{galeski2022signatures}
Galeski S, Legg H, Wawrzy{\'n}czak R, F{\"o}rster T, Zherlitsyn S, Gorbunov D,
  Uhlarz M, Lozano P, Li Q, Gu G {\em et~al.\/} 2022 {\em Nat. Commun.\/} {\bf
  13} 7418 \urlprefix\url{https://www.nature.com/articles/s41467-022-35106-7}

\bibitem{piva2024importance}
Piva M~M, Wawrzy\ifmmode~\acute{n}\else \'{n}\fi{}czak R, Kumar N, Kutelak L~O,
  Lombardi G~A, dos Reis R~D, Felser C and Nicklas M 2024 {\em Phys. Rev.
  Mater.\/} {\bf 8}(4) L041202
  \urlprefix\url{https://link.aps.org/doi/10.1103/PhysRevMaterials.8.L041202}

\bibitem{tang2019three}
Tang F, Ren Y, Wang P, Zhong R, Schneeloch J, Yang S~A, Yang K, Lee P~A, Gu G,
  Qiao Z {\em et~al.\/} 2019 {\em Nature\/} {\bf 569} 537--541
  \urlprefix\url{https://www.nature.com/articles/s41586-019-1180-9}

\bibitem{galeski2021origin}
Galeski S, Ehmcke T, Wawrzy{\'n}czak R, Lozano P~M, Cho K, Sharma A, Das S,
  K{\"u}ster F, Sessi P, Brando M {\em et~al.\/} 2021 {\em Nat. Comm.\/} {\bf
  12} 3197 \urlprefix\url{https://www.nature.com/articles/s41467-021-23435-y}

\bibitem{ramshaw2018quantum}
Ramshaw B, Modic K~A, Shekhter A, Zhang Y, Kim E~A, Moll P~J, Bachmann M~D,
  Chan M, Betts J, Balakirev F {\em et~al.\/} 2018 {\em Nat. Commun.\/} {\bf 9}
  2217 \urlprefix\url{https://www.nature.com/articles/s41467-018-04542-9}

\bibitem{rosa2019enhanced}
Rosa P~F~S, Thomas S~M, Balakirev F~F, Bauer E~D, Fernandes R~M, Thompson J~D,
  Ronning F and Jaime M 2019 {\em Phys. Rev. Lett.\/} {\bf 122}(1) 016402
  \urlprefix\url{https://link.aps.org/doi/10.1103/PhysRevLett.122.016402}

\bibitem{lesseux2020orbitally}
Lesseux G~G, Sakai H, Hattori T, Tokunaga Y, Kambe S, Kuhns P~L, Reyes A~P,
  Thompson J~D, Pagliuso P~G and Urbano R~R 2020 {\em Phys. Rev. B\/} {\bf
  101}(16) 165111
  \urlprefix\url{https://link.aps.org/doi/10.1103/PhysRevB.101.165111}

\bibitem{souza2022magnetic}
Souza J~C 2022 {\em {Magnetic properties of nontrivial topological complex
  systems= Propriedades magn{\'e}ticas de sistemas complexos com topologia
  n{\~a}o trivial}\/} Ph.D. thesis Universidade Estadual de Campinas, Instituto
  de Física Gleb Wataghin, Campinas, SP
  \urlprefix\url{https://hdl.handle.net/20.500.12733/4556}

\bibitem{aoki1993transition}
Aoki H, Uji S, Albessard A~K and \ifmmode~\bar{O}\else \={O}\fi{}nuki Y 1993
  {\em Phys. Rev. Lett.\/} {\bf 71}(13) 2110--2113
  \urlprefix\url{https://link.aps.org/doi/10.1103/PhysRevLett.71.2110}

\bibitem{kozlova2005magnetic}
Kozlova N, Hagel J, Doerr M, Wosnitza J, Eckert D, M\"uller K~H, Schultz L,
  Opahle I, Elgazzar S, Richter M, Goll G, v~L\"ohneysen H, Zwicknagl G,
  Yoshino T and Takabatake T 2005 {\em Phys. Rev. Lett.\/} {\bf 95}(8) 086403
  \urlprefix\url{https://link.aps.org/doi/10.1103/PhysRevLett.95.086403}

\bibitem{bastien2016lifshitz}
Bastien G, Gourgout A, Aoki D, Pourret A, Sheikin I, Seyfarth G, Flouquet J and
  Knebel G 2016 {\em Phys. Rev. Lett.\/} {\bf 117}(20) 206401
  \urlprefix\url{https://link.aps.org/doi/10.1103/PhysRevLett.117.206401}

\bibitem{pfau2017cascade}
Pfau H, Daou R, Friedemann S, Karbassi S, Ghannadzadeh S, K\"uchler R, Hamann
  S, Steppke A, Sun D, K\"onig M, Mackenzie A~P, Kliemt K, Krellner C and
  Brando M 2017 {\em Phys. Rev. Lett.\/} {\bf 119}(12) 126402
  \urlprefix\url{https://link.aps.org/doi/10.1103/PhysRevLett.119.126402}

\bibitem{pfau2013interplay}
Pfau H, Daou R, Lausberg S, Naren H~R, Brando M, Friedemann S, Wirth S,
  Westerkamp T, Stockert U, Gegenwart P, Krellner C, Geibel C, Zwicknagl G and
  Steglich F 2013 {\em Phys. Rev. Lett.\/} {\bf 110}(25) 256403
  \urlprefix\url{https://link.aps.org/doi/10.1103/PhysRevLett.110.256403}

\end{thebibliography}

\end{document}